\documentclass[notitlepage, superscriptaddress, showpacs, nobalancelastpage,longbibliography, twocolumn, aps, prl]{revtex4-2}
\usepackage[english]{babel}
\RequirePackage[T1]{fontenc}
\RequirePackage{times} 

\usepackage{siunitx} 
\usepackage[italicdiff]{physics}
\usepackage{amsfonts}
\usepackage{lipsum}
\usepackage{multirow}
\usepackage{subfigure}
\usepackage{amsmath}
\usepackage{amssymb}
\usepackage{graphicx, epstopdf}
\usepackage{xspace}
\usepackage{fullwidth}
\usepackage{array}
\usepackage[hidelinks]{hyperref}
\usepackage[dvipsnames]{xcolor}
\usepackage{my_symbols}
\usepackage{CJK}
\usepackage{bm}
\usepackage{verbatim}
\usepackage{tikz}
\usepackage{comment}
\usepackage{marginnote}
\usepackage[textwidth=1.5cm]{todonotes}
\usepackage[normalem]{ulem}
\usepackage{soul}

\sethlcolor{green}

\newcommand{\ii}{\mathrm{i}}
{\par}

\newcounter{mycomment}

\excludecomment{comment}  

\begin{document}

\begin{CJK*}{UTF8}{gbsn} 
\title{Alternating Spintronics: Capacitive Behavior of Spin Valves and Resonator Applications}

\author{Yunwen Liu}
\affiliation{Department of Physics and State Key Laboratory of Surface Physics, Fudan University, Shanghai 200433, China}
\author{Jiang Xiao (萧江)}
\affiliation{Department of Physics and State Key Laboratory of Surface Physics, Fudan University, Shanghai 200433, China}
\affiliation{Institute for Nanoelectronics Devices and Quantum Computing, Fudan University, Shanghai 200433, China}
\affiliation{Shanghai Research Center for Quantum Sciences, Shanghai 201315, China}
\affiliation{Hefei National Laboratory, Hefei 230088, China}

\begin{abstract}
This study explores the time-dependent spin transport phenomena in magnetic heterostructures under alternating currents (AC), advancing the relatively underdeveloped field of alternating spintronics. Employing a time-dependent spin diffusion model, we show that the interplay of AC frequencies and spin relaxation times reveals significant differences in spin accumulation patterns compared to conventional direct current (DC) scenarios. Of particular interest is the emergence of capacitive-like impedance in a spin valve under AC conditions, which is especially pronounced in antiparallel spin configurations. 
These findings open up possibilities for developing high-frequency spintronic devices, including the proposed "spin resonator", which functions like a standard LC resonator but without a traditional capacitor.
\end{abstract}
\maketitle
\end{CJK*}

\emph{Introduction -- } 
More than a century ago, the scientific community witnessed a fierce rivalry between Thomas Edison and Nikola Tesla over the best means for electrical power delivery \cite{jonnes2003empires}. Edison championed direct current (DC), advocating for its safety and simplicity, while Tesla promoted alternating current (AC) for its efficiency in long-distance transmission. Ultimately, AC became the dominant method for electrical grids due to its advantages in power transmission and distribution. Nevertheless, both AC and DC have found their unique roles, each suited to specific applications ranging from power supply systems to telecommunications and information processing.

The field of spintronics seeks to utilize not only the charge of the electron but also its intrinsic spin and associated magnetic moment \cite{maekawa_concepts_2006}. Spintronics holds significant promise for transforming conventional electronics, potentially enabling devices with superior speed, energy efficiency, and new functionalities. In recent years, the applications of spintronics have evolved from information storage \cite{sverdlov_nonvolatile_2017,yang_2d_2022} towards unconventional computing such as the neuromorphic \cite{grollier_spintronic_2016}, probabilistic \cite{chowdhury_full-stack_2023}, or quantum computing \cite{he_topological_2019,qin_memristor_2023,hu_single_2025}.

Interestingly, most spintronic research has stayed almost exclusively on the DC regime, such as the Giant/Tunneling Magnetoresistance (GMR, TMR) \cite{baibich_giant_1988}, spin-transfer torque \cite{slonczewski_current-driven_1996,berger_emission_1996}, spin-orbit torque, spin Hall effect \cite{sinova_universal_2004}, and even spin pumping \cite{tserkovnyak_enhanced_2002} which involves magnetization dynamics, \etc. 
In contrast, the concept of alternating spin currents - where the flowing direction of spin currents or equivalently the spin polarization changes periodically - has received very little attention. The few exceptions include the study of AC spin pumping \cite{chiba_current_2014,li_ac_2016,kapelrud_pumping_2017}, and a subfield of spintronics on the THz generation in ferromagnet-noble metal bilayers \cite{kampfrath_THz_2013}, in which cases alternating spin currents are generated and detected. This predominant focus on the direct spin current is mainly due to the fact that spin relaxation is extremely fast in comparison with the AC frequency, and therefore a varying current, if exists, can be considered as quasi-static in the viewpoint of spin relaxation, so there is no need to care about any time-dependent effects. However, as some new materials emerge with extremely long spin relaxation times, such as the 2D van der Waals materials \cite{droegeler_lifetime_2016,xu_spin_2020}, the alternating spin transport becomes relevant. In addition, inspired from the THz generation in ferromagnet-noble metals, we see that materials with short spin relaxation time can be used for high frequency applications \cite{lu_THz_2024,bhide_6g_2024}. Therefore, understanding and utilizing alternating spin currents could eventually lead to new spintronic device concepts and new paradigms in future high speed communication and processing technologies.

In this paper, we explore the dynamic behavior of spin transport in magnetic heterostructures exposed to alternating currents. Using a time-dependent spin diffusion model, we investigate how spin accumulation develops resistive and capacitive responses when an AC current is applied. The results show notable differences from conventional DC spintronic systems, such as the appearance of a capacitive-like impedance in spin valves. These findings suggest promising possibilities for high-frequency applications, such as the development of a "spin resonator".


\emph{Time-dependent spin diffusion -- }
We consider a one-dimensional model with alternating charge current $j(t)$ driving through a magnetic heterostructure consisting of one or more normal and ferromagnetic metallic layers, such as the normal-ferromagnet junction or ferromagnet-normal-ferromagnet spin valve. The spin accumulation $\bmu(x,t)$ developed in NM and FM becomes time-dependent when the driving charge current $j(t)$ is an alternating current or the magnetization $\mb(t)$ is in a dynamical precession. The equation governing the magnetization dynamics (in FM) and spin accumulation (in both FM and NM) are the Landau-Lifshitz-Gilbert (LLG) equation and the spin diffusion equation, respectively: 
\begin{subequations}
    \label{eqn:LLG-sd}
    \begin{align}
    \pdv{\mb(t)}{t} &= -\gamma \mb \times \bH_\text{eff} + \alpha \mb \times \dot{\mb}, \label{eqn:LLG} \\ 
    \pdv{\bmu^i(x,t)}{t} &= D^i \pdv[2]{\bmu^i}{x} - \frac{\bmu^i}{\tau_s^i} + \fb^i(x,t), \label{eqn:SD}
    \end{align}
\end{subequations}
where $i = $ N or F labels the NM or FM layer, $\bH_\text{eff}$ the effective magnetic field, $\alpha$ are the Gilbert damping constant in FM, $\gamma$ is the gyromagnetic ratio, $\tau_s^i$ the spin relaxation time. And $\fb^i(x,t)$ is the local spin generation rate. The diffusion coefficient $D^i$, according to Nernst-Einstein relation \cite{stiles_phenomenological_2004}, is proportional to the conductivity: $D^i = (\partial n/\partial \mu)^{-1} (\sigma^i/e^2) 
$.
In this work, we set $\fb = 0$. However, for situations such as terahertz generation in NF bilayers with femtosecond laser \cite{battiato_superdiffusion_2010,kampfrath_THz_2013}, such local spin generation $\fb$ can be modeled by a pulse with a certain spatial profile.
We adopt the two-carrier model. The spin-dependent conductivities in FM are $\sigma_{\up, \dn}^\ssf{F}$ for spin up and down channels. The corresponding spin polarization is $\eta = (\sigma_\up^\ssf{F} - \sigma_\dn^\ssf{F})/\sigma^\ssf{F}$ with total conductivity $\sigma^\ssf{F} = \sigma_\up^\ssf{F}+\sigma_\dn^\ssf{F}$.  
The overall diffusion coefficient in F
$D^\ssf{F} = (D^\ssf{F}_\up\sigma^\ssf{F}_\dn+D^\ssf{F}_\dn\sigma^\ssf{F}_\up)/\sigma^\ssf{F}
=(1-\eta^2)\sigma^\ssf{F}/2$ \cite{van_son_boundary_1987,stiles_phenomenological_2004}. Similar but simpler definitions hold for NM with $\sigma_\up^\ssf{N} = \sigma_\dn^\ssf{N}$ and $\eta = 0$.

In FM, we assume that the spin accumulation is collinear with the local magnetization: $\bmu^\ssf{F}(x) = \mu^\ssf{F}\mb(x)$ \cite{stiles_noncollinear_2002, stiles_anatomy_2002,stiles_phenomenological_2004}. And in NM, the spin accumulation can have both longitudinal and transverse components: $\bmu^\ssf{N} = \mu_\|^\ssf{N}\mb + \bmu_\perp^\ssf{N}$ with $\mu_\|^\ssf{N} = \bmu^\ssf{N}\cdot\mb$ and $\bmu^\ssf{N}_\perp \equiv \bmu^\ssf{N} - \mu_\|^\ssf{N}\mb$.
At the NM-FM interfaces, we require the continuity of the longitudinal spin accumulation $\mu_\|^\ssf{N} = \mu^\ssf{F}$, and the longitudinal and transverse spin current (torque) across the interface \cite{stiles_phenomenological_2004}:
\begin{subequations}
    \label{eqn:bc}
\begin{align}
    - D^\ssf{N} \pdv{\mu^\ssf{N}_\|}{x}  &= \eta e j(t) - D^\ssf{F} \pdv{\mu^\ssf{F}}{x}, \label{eqn:bc_long} \\
    - D^\ssf{N} \pdv{\bmu^\ssf{N}_\perp}{x} 
    &= -A_\ssf{ex} \pdv{\mb}{x} 
    = g~\mb\times\qty(\bmu^\ssf{N}_\perp\times\mb - \hbar~\dot{\mb}). \label{eqn:bc_trans}
\end{align}
\end{subequations}

In this model, three distinct timescales are relevant: the magnetization dynamics timescale ($\tau_m$), determined by the magnetic precession frequency; the spin relaxation time ($\tau_s$), described by the spin diffusion equation; and the timescale of the alternating current ($\tau_c$), which depends on the external AC driving frequency. Previous studies have typically assumed a strict separation of these timescales, with spin relaxation being the fastest and the driving current the slowest process (\ie, $\tau_s \ll \tau_m \ll \tau_c$). However, this work explores the less-studied regime where the spin relaxation time and the current alternation time are comparable ($\tau_s \sim \tau_c$), while the magnetization remains effectively static. In most conventional metals, spin relaxation occurs on the order of picoseconds, corresponding to THz frequencies, which makes matching these timescales with current alternation challenging for standard electronics. Nevertheless, in spintronic devices - such as those utilizing femtosecond laser-induced THz emission in ferromagnet-platinum bilayers \cite{kampfrath_THz_2013} - these high frequencies are accessible. Furthermore, in materials with weak spin-orbit coupling, especially low-dimensional systems like graphene, spin relaxation can be orders of magnitude slower, extending to nanoseconds or even microseconds \cite{droegeler_lifetime_2016,xu_spin_2020,zhou_cdx_2023}. In such cases, current modulations in the GHz or MHz range suffice to reach the condition $\tau_c \sim \tau_s$.

In such circumstances, the time dependence of the system is purely due to the alternating current driven into the system, and we may neglect the magnetization dynamics or the LLG equation in \Eq{eqn:LLG} and the boundary condition for the transverse spin current in \Eq{eqn:bc_trans}. Consequently, we only need to solve for the spin diffusion equation \Eq{eqn:SD} in NM and FM with the continuity boundary condition for the longitudinal spin accumulation and spin current to find the longitudinal spin accumulation $\bmu(x,t)$. Because of the polarization difference between two materials, an abrupt change in the electric chemical potential appears across the interface, resulting in an interfacial impedance of the FN-NM interface (at $x = 0$) \cite{van_son_boundary_1987}:
\begin{equation}
    \label{eqn:Z}
    Z = \frac{\eta\mu^\ssf{F}(0,t)}{ej(t)A}
      = \frac{\eta\mu_\|^\ssf{N}(0,t)}{ej(t)A},
\end{equation}
where $A$ is the cross-section area of the junction. The second equality is because $\mu^\ssf{F} = \mu_\|^\ssf{N}$ across the FN interface.

For the collinear magnetic heterostructures focused in this work, the spin quantization axis is along the magnetization direction in FM, thus $\bmu^i = \mu^i\mb$. 
For the driving charge current $j(t) = j e^{\ii\omega t}$, we may assume  $\mu^i(x,t) = \tilde{\mu}^i(x) e^{\ii\omega t}$ in both N and F, where $\Re(\tilde{\mu})$ and $\Im(\tilde{\mu})$ represent the in-phase and out-of-phase components of the spin accumulation induced by the driving current. 
The spin diffusion equation becomes 
\begin{equation}
    \ii\omega \tau_s^i~\tilde{\mu}^i(x) 
    = {l_s^i}^2 \pdv[2]{x}\tilde{\mu}^i(x) - \tilde{\mu}^i(x)
\end{equation}
with spin diffusion length $l_s^i = \sqrt{D^i\tau_s^i}$.
The solution is
\begin{equation}
    \label{eqn:mukappa}
    \tilde{\mu}^i(x) = A_i e^{\kappa_i x} + B_i e^{-\kappa_i x}
    ~~\text{with}~~ 
    \kappa_i(\omega) = \frac{\sqrt{1+\ii\omega \tau_s^i}}{l_s^i}.
\end{equation}
What's crucial here is the frequency-dependent and complex-valued $\kappa_i(\omega)$, which distinguishes itself from the real valued one in the time-independent spin diffusion models assumed in most previous studies.

\emph{Spin Valve -- }
Spin valve, a common magnetic heterostructure, comprises two identical ferromagnetic (FM) layers separated by a non-magnetic (NM) spacer (of thickness $d$). The spin accumulation in the three layers can be expressed as
\begin{align}
    \mu^\ssf{F}(x) &= A_\ssf{F} e^{\kappa_\ssf{F} x} &\qfor& x \le 0, \nn
    \mu^\ssf{N}(x) &= A_\ssf{N} e^{\kappa_\ssf{N} x} + B_\ssf{N} e^{-\kappa_\ssf{N} x} &\qfor& 0 < x < d, 
    \label{eqn:muFNF} \\
    \mu^\ssf{F}(x) &= B_\ssf{F} e^{-\kappa_\ssf{F} (x-d)} &\qfor& x \ge d. \nonumber
\end{align}
The continuity of spin accumulation requires
\begin{equation}
    A_\ssf{F} = A_\ssf{N} + B_\ssf{N} \qand 
    B_\ssf{F} = A_\ssf{N} e^{\kappa_\ssf{N} d} + B_\ssf{N} e^{-\kappa_\ssf{N} d}.
\label{eqn:BC1}
\end{equation}
The continuity of the longitudinal spin current requires
\begin{subequations}
\label{eqn:BC2}
\begin{align}
    \eta e j - D^\ssf{F} \kappa_\ssf{F} A_\ssf{F} &= - D^\ssf{N} \kappa_\ssf{N} (A_\ssf{N} - B_\ssf{N}), \\
    \eta' e j + D^\ssf{F} \kappa_\ssf{F} B_\ssf{F} &= - D^\ssf{N} \kappa_\ssf{N} (A_\ssf{N} e^{\kappa_\ssf{N} d} - B_\ssf{N} e^{-\kappa_\ssf{N} d}),
\end{align}
\end{subequations}
where $\eta, \eta'$ are the spin current polarization in the two FM layers, and $\eta' = \pm \eta$ if the magnetization in the two FMs (of the same material) are (anti-)parallel.

\begin{figure}[t]
\includegraphics[width=\columnwidth]{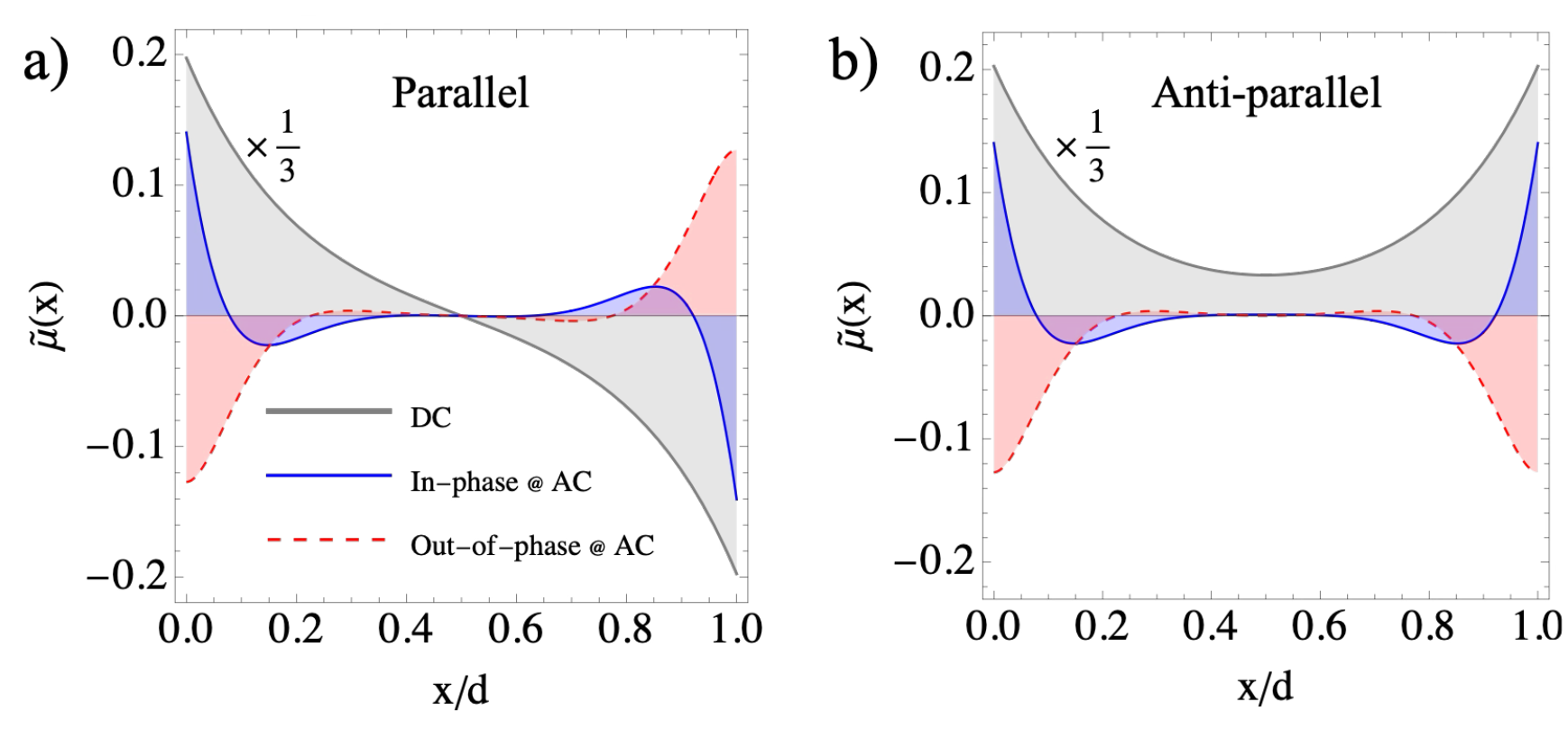}
\caption{Spin accumulation within the spacer in a spin valve for (anti-)parallel configurations. The solid gray (scaled down by $1/3$) curve is the (in-phase) spin accumulation under DC driving. The solid (blue) and dashed (red) curves are the real (in-phase) and imaginary (out-of-phase) part of the spin accumulation under AC driving at $\omega\tau_s = 10$. Spin diffusion lengths: $l_s^\ssf{N}/d = 0.2, l_s^\ssf{F} \ra 0$.}
\label{fig:ux}
\end{figure}

By solving \Eqss{eqn:muFNF}{eqn:BC2}, we find the distribution of the complex spin accumulation across the structure. \Figure{fig:ux} shows the comparison of the spin accumulation in the normal metal spacer under DC and AC driving: The gray curves are the spin accumulation under DC driving ($\omega = 0$), and the solid-blue and dashed-red ones are the in-phase and out-of-phase spin accumulation under AC driving ($\omega\tau_s = 10$). It can be seen that the out-of-phase spin accumulation can be comparable to the in-phase component under the alternating driving current for both parallel and antiparallel configurations. Such out-of-phase component will eventually lead to a capacitive-like behavior in the boundary impedance. 

The impedance due to the spin accumulation is given by the total chemical potential drop at the two interfaces ($x =0, d$):
\begin{align}
    \label{eqn:ZFNF}
    &Z(\omega) = Z_0 + Z_d 
    = \frac{\eta\mu^\ssf{F}(0)-\eta'\mu^\ssf{F}(d)}{ejA} 
    = \frac{\eta A_\ssf{F}- \eta' B_\ssf{F}}{ejA} \nn
    &= \begin{cases}
    \frac{2\eta^2/A}{D^\ssf{F}\kappa_\ssf{F} + D^\ssf{N}\kappa_\ssf{N} \coth\frac{\kappa_\ssf{N}d}{2}} 
    \ra \frac{2\eta^2}{D^\ssf{N}\kappa_\ssf{N}A} \tanh\frac{\kappa_\ssf{N}d}{2} 
    \quad \text{(P)} \\
    \frac{2\eta^2/A}{D^\ssf{F}\kappa_\ssf{F} + D^\ssf{N}\kappa_\ssf{N} \tanh\frac{\kappa_\ssf{N}d}{2}} 
    \ra
    \frac{2\eta^2}{D^\ssf{N}\kappa_\ssf{N}A} \coth\frac{\kappa_\ssf{N}d}{2}
    \quad \text{(AP)} 
    \end{cases}
\end{align}
where the frequency dependence of the impedance is carried through the complex value of $\kappa_\ssf{F, N}(\omega)$ in \Eq{eqn:mukappa}. The limiting expressions are for $(1-\eta^2)\sigma^\ssf{F}\propto D^\ssf{F}\ra 0$, \ie the spin diffusion in F is very short in comparison with that in N. 

\begin{figure}[t] 
    \includegraphics[width=\columnwidth]{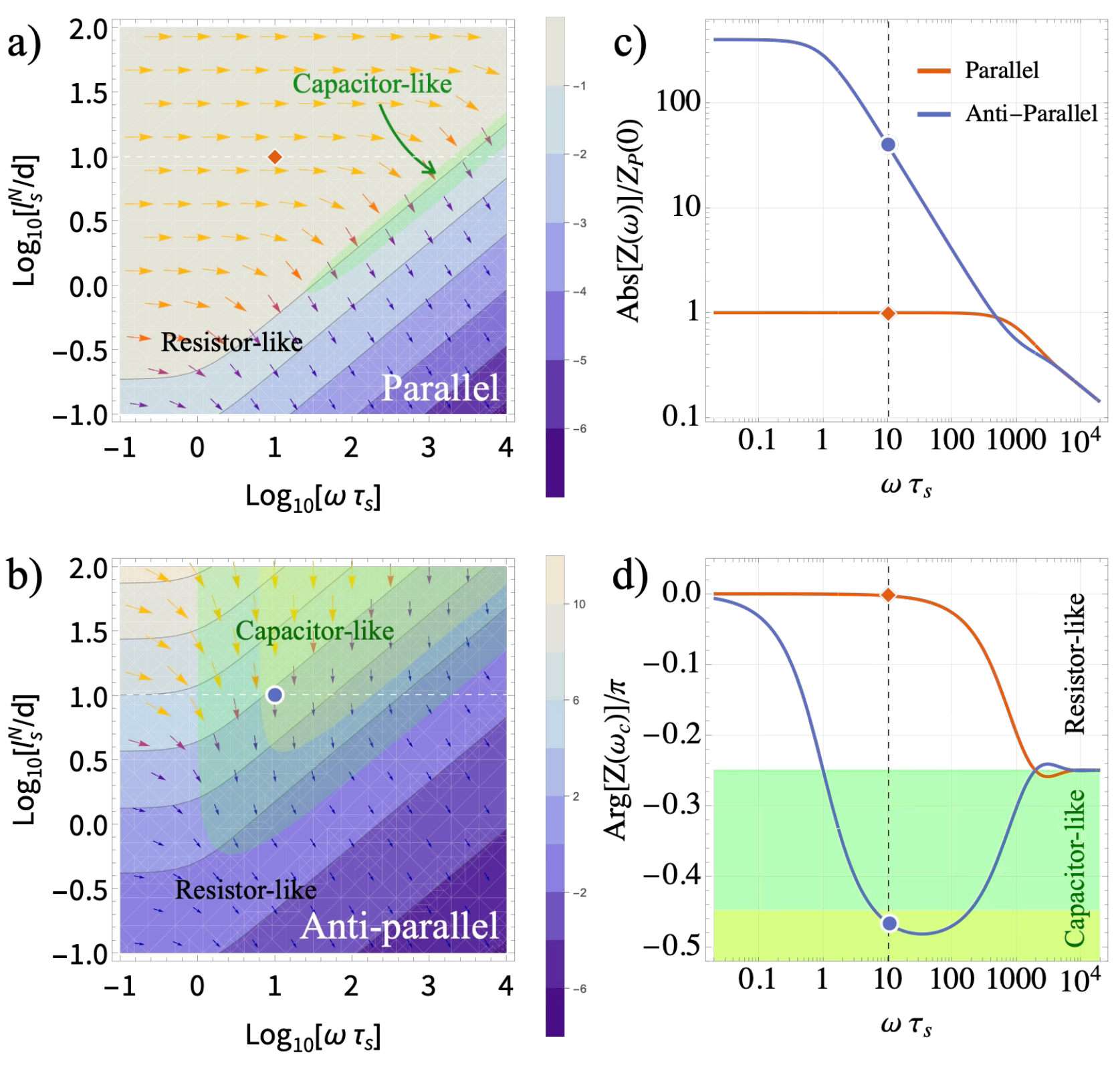} 
    \caption{
    a, b) The magnitude (colormap, $\log(\abs{Z})$) and the phase (arrows, $\arg{Z}$) of $Z$ for spin valve in (anti)parallel configuration. The green (yellow) shaded area indicates the region with $\text{arg}(Z) < -\pi/4$ ($-\pi/2\times 90\%$). 
    c, d) The magnitude and the argument of $Z$ as function of driving frequency $\omega_c$ for $l_s^\ssf{N}/d = 10$.
    The vertical dashed line and the points mark an example working point at $\omega_c\tau_s = 10$.
    }
    \label{fig:FNF}
\end{figure}


\Figure{fig:FNF}(a, b) shows the impedance $Z(\omega)$ for the parallel and antiparallel configurations, in the limit of $D^\ssf{F} \to 0$, against the driving frequency $\omega$ and the normalized spin diffusion length in the spacer $l_s^\ssf{N}/d$.
The color scale represents the magnitude $\abs{Z}$, while the direction of the arrows indicate the phase $\arg{Z}$. The magnitude $\abs{Z}$ decreases with increasing frequency, consistent with the expectation that higher frequencies allow less time for spin accumulation to occur. The phase $\arg{Z}$ distinguishes the resistive and capacitive contributions. Arrows pointing to the right correspond to a resistive regime ($\arg{Z} \sim 0$), while those pointing downward indicate a capacitive regime ($\arg{Z} \sim -\pi/2$). The green-shaded regions highlight the capacitor-like region defined by $\arg{Z} \leq -\pi/4$. This capacitor-like region is significantly more pronounced in the antiparallel configuration, particularly under conditions where the driving frequency is faster than the spin relaxation rate ($\omega \tau_s > 1$) and the spin diffusion length exceeds the spacer thickness ($l_s^\ssf{N} > d$). Interestingly, while the parallel configuration only marginally enters the capacitor-like regime (with $\arg{Z} \gtrsim -\pi/4$), the antiparallel configuration not only exhibits a much larger capacitor-like regime, but can also become nearly purely capacitive: in the darker green highlighted region in \Figure{fig:FNF}(b) the phase approaches $-\pi/2$ with $\arg{Z} < -\pi/2\times 90\%$. This behavior is further elucidated in \Figure{fig:FNF}(c, d),
which provides a line-cut plot for $l_s^\ssf{N}/d = 10$, corresponding to the top edge of \Figure{fig:FNF}(a, b). 
The magnitude $\abs{Z}$ for the antiparallel configuration is much larger than the parallel configuration, simply because the magnitude of the spin accumulation is larger in the antiparallel case. The phase $\arg{Z}$ reveals distinct behaviors. At high driving frequencies ($\omega \tau_s > 1000$), the phases for both configurations approaches $-\pi/4$, indicating a balance between resistive and capacitive contributions. However, at the intermediate frequency ($\omega \tau_s \sim 10 - 100$), the phase for the antiparallel configuration approaches $-\pi/2$, signifying a purely capacitive response.  
This implies that the FNF junction in antiparallel configuration can function as a pure capacitor. 
Furthermore, the magnitude $\abs{Z}$ of the capacitor is approximately $10\%$ of the DC magnetoresistance: $|Z_\ssf{AP}(\omega \tau_s \sim 10)/Z_\ssf{AP}(0)| \sim 0.1$. This means that, in the antiparallel configuration, a purely capacitive impedance is not only achievable, but also of appropriate magnitude.

\begin{figure}[t]
\includegraphics[width=\columnwidth]{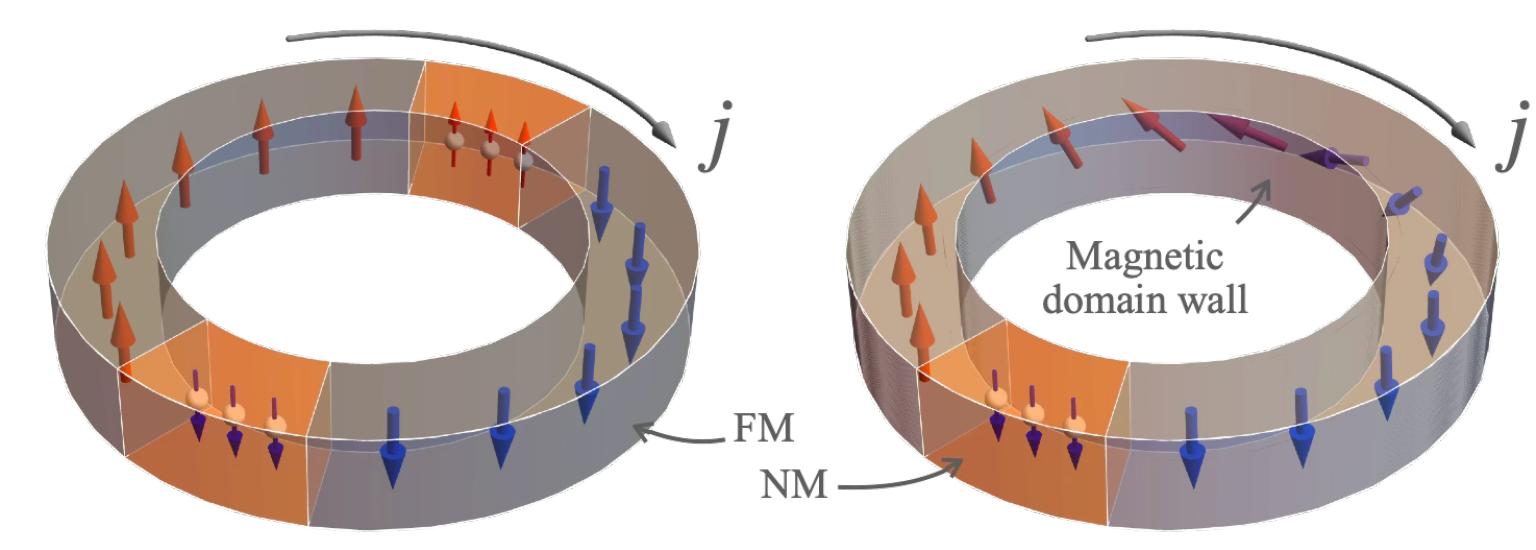}
\caption{Two realizations of spin resonator based on two interconnecting FNF junctions (left) and a single FNF junction formed by joining the two ends of a magnetic domain wall (right). Arrows in the FM segments indicate the magnetization direction. Arrows in the NM segment indicate the polarization direction of the spin accumulation induced by the instantaneous current $j$.}
\label{fig:ring}
\end{figure}

\emph{Spin Resonator -- } 
Based on the predicted capacitive behavior of spin valves, we propose a new spintronic device: the spin resonator. 
This resonator can be realized in two ways as depicted in \Figure{fig:ring}: a closed-loop structure of two interconnected spin valves with antiparallel magnetization, or a single spin valve formed by joining the ends of a magnetic domain wall. In both realizations, the antiparallel FNF junction acts as a capacitor, while the ring structure provides the inductive component, creating an LC resonator equivalent.

Since the FNF junction is capacitor-like only within a certain frequency range ($\omega \tau_s^\ssf{N} \sim 10$), we shall require the resonant frequency of the spin resonator $\omega_0 = 1/\sqrt{LC}$ falls in this frequency range: $\omega_0 \sim \omega \sim 10/\tau_s^\ssf{N}$. This condition puts a constraint on the size of the ring, which we estimate as the following: The inductance of the structure is proportional to the radius of the ring $r$: $L \sim \mu_0 r$. The magnitude of the nearly purely capacitor-like impedance is about $10\%$ of the DC interfacial resistance at $l_s^\ssf{N}/d \sim 10$ (see the point in \Figure{fig:FNF}(c)): $C \sim [\omega_0 Z_\ssf{AP}(0)\times 10\%]^{-1}$. By requiring the value of $L$ and $C$ estimated above result in a self-consistent resonant frequency $\omega_0$, we find that the radius of the ring (see Appendix)
\begin{equation}
    r \sim \frac{\rho^\ssf{N} d\tau_s^\ssf{N}}{A \mu_0} \simeq \frac{\rho^\ssf{N}l_s^\ssf{N}}{A\mu_0\omega_0},
    \label{eqn:r}
\end{equation}
For the most common metallic spacer of Cu ($\rho^\ssf{Cu} = \SI{21}{\Omega\cdot nm}, l_s^\ssf{Cu} = \SI{500}{nm}, \tau_s^\ssf{Cu} \simeq \SI{0.02}{ns}, A = 100\times\SI{10}{nm^2}$ \cite{,stiles_phenomenological_2004,bass_spin-diffusion_2007}), the ring radius of $r \simeq \SI{10}{\mu m}$ may realize consistently the resonant frequency of $\omega_0 \simeq \SI{1}{THz}$. 

The quality factor $Q$ for a resonator with resistance $R$ is given by $Q = (\omega_0CR)^{-1}$.
The resistance of the spin resonator includes the bulk resistance from the FM and NM layers and the resistive part of the interfacial impedance: $R \sim R_\ssf{bulk} +  \Re(Z_\ssf{AP})$. 
The capacitive part $1/\omega_0C=\Im(Z_\ssf{AP})$. Therefore, $Q = \abs{\Im(Z_\ssf{AP})}/[R_\ssf{bulk} + \Re(Z_\ssf{AP})] < \abs{\tan (\arg{Z})}$, which has a very high upper bound if the interfacial impedance dominates the resistance and the phase approaches $-\pi/2$ (purely capacitive). For the spin resonator with parameters given above for Cu, we estimate $Q \sim 10$. In comparison with the applications in GHz range, this quality factor may not be high, but good enough for many THz applications. 


The spin resonator offers unique tunability, as its resonant properties can be switched on or off by altering the magnetic structure within the FM layers. When the FM layers are in an antiparallel alignment, or if a domain wall exists, the FNF junction acts as a resonator due to enhanced capacitive behavior. In contrast, a parallel alignment, or the absence of a domain wall, greatly diminishes capacitance and suppresses resonance. This feature allows the device's resonance to be actively controlled in real time through external means, such as applying a magnetic field or using spin-transfer torque.

\emph{Discussion \& Conclusion -- }
The configuration dependence of the capacitance in a spin valve makes it possible to use capacitance as a way to detect magnetic states. In steady-state measurements, a small AC signal is applied, and the resulting phase shift varies between parallel and antiparallel configurations. In transient conditions, sending a current pulse causes stronger discharging in the antiparallel state compared to the parallel state. These behaviors highlight capacitive readout as an alternative to conventional magnetoresistance-based techniques for detecting magnetic states.


Two-dimensional materials, exemplified by graphene owing to its long spin relaxation time, offer a promising platform for exploring the principles of alternating spintronics. However, the pronounced resistive characteristics frequently observed in such materials \cite{novoselov_2d_2004,xue_graphene_2021,cultrera_iec_2024} may pose some challenges for alternating spintronic applications.

This work assumed that the magnetization of the ferromagnetic element is static. However, if the magnetization dynamics are included, the spin resonator can exhibit richer behaviors such as non-linearity. The interaction between the resonator's resonance and magnetization dynamics may lead to more complex responses and enable more unique functionalities beyond the static case. 

In summary, our investigation of time-dependent spin transport in magnetic heterostructures under alternating current has uncovered significant distinctions between AC and conventional DC spintronics. We demonstrate that a spin valve in the antiparallel configuration can display a prominent capacitive impedance, thereby enabling capacitive readout of the magnetic state. This finding paves the way for a new class of AC spintronic devices such as the spin resonator, a spin analog of LC oscillator, which holds promise for operation at terahertz frequencies.



\emph{Acknowledgements. } 
This work was supported by 
National Natural Science Foundation of China (Grants No. 12474110),
the National Key Research and Development Program of China (Grant No. 2022YFA1403300),
the Innovation Program for Quantum Science and Technology (Grant No.2024ZD0300103),
and Shanghai Municipal Science and Technology Major Project (Grant No.2019SHZDZX01).

\bibliography{all,MTJ}

\begin{thebibliography}{32}%
\makeatletter
\providecommand \@ifxundefined [1]{%
 \@ifx{#1\undefined}
}%
\providecommand \@ifnum [1]{%
 \ifnum #1\expandafter \@firstoftwo
 \else \expandafter \@secondoftwo
 \fi
}%
\providecommand \@ifx [1]{%
 \ifx #1\expandafter \@firstoftwo
 \else \expandafter \@secondoftwo
 \fi
}%
\providecommand \natexlab [1]{#1}%
\providecommand \enquote  [1]{``#1''}%
\providecommand \bibnamefont  [1]{#1}%
\providecommand \bibfnamefont [1]{#1}%
\providecommand \citenamefont [1]{#1}%
\providecommand \href@noop [0]{\@secondoftwo}%
\providecommand \href [0]{\begingroup \@sanitize@url \@href}%
\providecommand \@href[1]{\@@startlink{#1}\@@href}%
\providecommand \@@href[1]{\endgroup#1\@@endlink}%
\providecommand \@sanitize@url [0]{\catcode `\\12\catcode `\$12\catcode `\&12\catcode `\#12\catcode `\^12\catcode `\_12\catcode `\%12\relax}%
\providecommand \@@startlink[1]{}%
\providecommand \@@endlink[0]{}%
\providecommand \url  [0]{\begingroup\@sanitize@url \@url }%
\providecommand \@url [1]{\endgroup\@href {#1}{\urlprefix }}%
\providecommand \urlprefix  [0]{URL }%
\providecommand \Eprint [0]{\href }%
\providecommand \doibase [0]{https://doi.org/}%
\providecommand \selectlanguage [0]{\@gobble}%
\providecommand \bibinfo  [0]{\@secondoftwo}%
\providecommand \bibfield  [0]{\@secondoftwo}%
\providecommand \translation [1]{[#1]}%
\providecommand \BibitemOpen [0]{}%
\providecommand \bibitemStop [0]{}%
\providecommand \bibitemNoStop [0]{.\EOS\space}%
\providecommand \EOS [0]{\spacefactor3000\relax}%
\providecommand \BibitemShut  [1]{\csname bibitem#1\endcsname}%
\let\auto@bib@innerbib\@empty
\bibitem [{\citenamefont {Jonnes}(2003)}]{jonnes2003empires}%
  \BibitemOpen
  \bibfield  {author} {\bibinfo {author} {\bibfnamefont {J.}~\bibnamefont {Jonnes}},\ }\href@noop {} {\emph {\bibinfo {title} {Empires of Light: Edison, Tesla, Westinghouse, and the Race to Electrify the World}}}\ (\bibinfo  {publisher} {Random House},\ \bibinfo {address} {New York},\ \bibinfo {year} {2003})\ pp.\ \bibinfo {pages} {3--75}\BibitemShut {NoStop}%
\bibitem [{\citenamefont {Maekawa}(2006)}]{maekawa_concepts_2006}%
  \BibitemOpen
  \bibfield  {author} {\bibinfo {author} {\bibfnamefont {S.}~\bibnamefont {Maekawa}},\ }\href@noop {} {\emph {\bibinfo {title} {Concepts in spin electronics}}}\ (\bibinfo  {publisher} {Oxford University Press},\ \bibinfo {year} {2006})\BibitemShut {NoStop}%
\bibitem [{\citenamefont {Sverdlov}\ \emph {et~al.}(2017)\citenamefont {Sverdlov}, \citenamefont {Weinbub},\ and\ \citenamefont {Selberherr}}]{sverdlov_nonvolatile_2017}%
  \BibitemOpen
  \bibfield  {author} {\bibinfo {author} {\bibfnamefont {V.}~\bibnamefont {Sverdlov}}, \bibinfo {author} {\bibfnamefont {J.}~\bibnamefont {Weinbub}},\ and\ \bibinfo {author} {\bibfnamefont {S.}~\bibnamefont {Selberherr}},\ }\bibfield  {title} {\bibinfo {title} {Spintronics as a non-volatile complement to modern microelectronics},\ }\href@noop {} {\bibfield  {journal} {\bibinfo  {journal} {Informacije Midem - Journal of Microelectronics, Electronic Components and Materials}\ }\textbf {\bibinfo {volume} {47}},\ \bibinfo {pages} {195} (\bibinfo {year} {2017})}\BibitemShut {NoStop}%
\bibitem [{\citenamefont {Yang}\ \emph {et~al.}(2022)\citenamefont {Yang}, \citenamefont {Valenzuela}, \citenamefont {Chshiev}, \citenamefont {Couet}, \citenamefont {Dieny}, \citenamefont {Dlubak}, \citenamefont {Fert}, \citenamefont {Garello}, \citenamefont {Jamet}, \citenamefont {Jeong}, \citenamefont {Lee}, \citenamefont {Lee}, \citenamefont {Martin}, \citenamefont {Kar}, \citenamefont {Seneor}, \citenamefont {Shin},\ and\ \citenamefont {Roche}}]{yang_2d_2022}%
  \BibitemOpen
  \bibfield  {author} {\bibinfo {author} {\bibfnamefont {H.}~\bibnamefont {Yang}}, \bibinfo {author} {\bibfnamefont {S.~O.}\ \bibnamefont {Valenzuela}}, \bibinfo {author} {\bibfnamefont {M.}~\bibnamefont {Chshiev}}, \bibinfo {author} {\bibfnamefont {S.}~\bibnamefont {Couet}}, \bibinfo {author} {\bibfnamefont {B.}~\bibnamefont {Dieny}}, \bibinfo {author} {\bibfnamefont {B.}~\bibnamefont {Dlubak}}, \bibinfo {author} {\bibfnamefont {A.}~\bibnamefont {Fert}}, \bibinfo {author} {\bibfnamefont {K.}~\bibnamefont {Garello}}, \bibinfo {author} {\bibfnamefont {M.}~\bibnamefont {Jamet}}, \bibinfo {author} {\bibfnamefont {D.-E.}\ \bibnamefont {Jeong}}, \bibinfo {author} {\bibfnamefont {K.}~\bibnamefont {Lee}}, \bibinfo {author} {\bibfnamefont {T.}~\bibnamefont {Lee}}, \bibinfo {author} {\bibfnamefont {M.-B.}\ \bibnamefont {Martin}}, \bibinfo {author} {\bibfnamefont {G.~S.}\ \bibnamefont {Kar}}, \bibinfo {author} {\bibfnamefont {P.}~\bibnamefont {Seneor}}, \bibinfo {author} {\bibfnamefont {H.-J.}\ \bibnamefont {Shin}},\ and\ \bibinfo {author} {\bibfnamefont {S.}~\bibnamefont {Roche}},\ }\bibfield  {title} {\bibinfo {title} {Two-dimensional materials prospects for non-volatile spintronic memories},\ }\href {https://doi.org/10.1038/s41586-022-04768-0} {\bibfield  {journal} {\bibinfo  {journal} {Nature}\ }\textbf {\bibinfo {volume} {606}},\ \bibinfo {pages} {663} (\bibinfo {year} {2022})}\BibitemShut {NoStop}%
\bibitem [{\citenamefont {Grollier}\ \emph {et~al.}(2016)\citenamefont {Grollier}, \citenamefont {Querlioz},\ and\ \citenamefont {Stiles}}]{grollier_spintronic_2016}%
  \BibitemOpen
  \bibfield  {author} {\bibinfo {author} {\bibfnamefont {J.}~\bibnamefont {Grollier}}, \bibinfo {author} {\bibfnamefont {D.}~\bibnamefont {Querlioz}},\ and\ \bibinfo {author} {\bibfnamefont {M.~D.}\ \bibnamefont {Stiles}},\ }\bibfield  {title} {\bibinfo {title} {Spintronic {Nanodevices} for {Bioinspired} {Computing}},\ }\href {https://doi.org/10.1109/JPROC.2016.2597152} {\bibfield  {journal} {\bibinfo  {journal} {Proceedings of the IEEE}\ }\textbf {\bibinfo {volume} {104}},\ \bibinfo {pages} {2024} (\bibinfo {year} {2016})},\ \bibinfo {note} {00006}\BibitemShut {NoStop}%
\bibitem [{\citenamefont {Chowdhury}\ \emph {et~al.}(2023)\citenamefont {Chowdhury}, \citenamefont {Grimaldi}, \citenamefont {Aadit}, \citenamefont {Niazi}, \citenamefont {Mohseni}, \citenamefont {Kanai}, \citenamefont {Ohno}, \citenamefont {Fukami}, \citenamefont {Theogarajan}, \citenamefont {Finocchio}, \citenamefont {Datta},\ and\ \citenamefont {Camsari}}]{chowdhury_full-stack_2023}%
  \BibitemOpen
  \bibfield  {author} {\bibinfo {author} {\bibfnamefont {S.}~\bibnamefont {Chowdhury}}, \bibinfo {author} {\bibfnamefont {A.}~\bibnamefont {Grimaldi}}, \bibinfo {author} {\bibfnamefont {N.~A.}\ \bibnamefont {Aadit}}, \bibinfo {author} {\bibfnamefont {S.}~\bibnamefont {Niazi}}, \bibinfo {author} {\bibfnamefont {M.}~\bibnamefont {Mohseni}}, \bibinfo {author} {\bibfnamefont {S.}~\bibnamefont {Kanai}}, \bibinfo {author} {\bibfnamefont {H.}~\bibnamefont {Ohno}}, \bibinfo {author} {\bibfnamefont {S.}~\bibnamefont {Fukami}}, \bibinfo {author} {\bibfnamefont {L.}~\bibnamefont {Theogarajan}}, \bibinfo {author} {\bibfnamefont {G.}~\bibnamefont {Finocchio}}, \bibinfo {author} {\bibfnamefont {S.}~\bibnamefont {Datta}},\ and\ \bibinfo {author} {\bibfnamefont {K.~Y.}\ \bibnamefont {Camsari}},\ }\bibfield  {title} {\bibinfo {title} {A full-stack view of probabilistic computing with p-bits: devices, architectures and algorithms},\ }\href {https://doi.org/10.1109/JXCDC.2023.3256981} {\bibfield  {journal} {\bibinfo  {journal} {IEEE Journal on Exploratory Solid-State Computational Devices and Circuits}\ ,\ \bibinfo {pages} {1}} (\bibinfo {year} {2023})},\ \bibinfo {note} {conference Name: IEEE Journal on Exploratory Solid-State Computational Devices and Circuits}\BibitemShut {NoStop}%
\bibitem [{\citenamefont {He}\ \emph {et~al.}(2019)\citenamefont {He}, \citenamefont {Sun},\ and\ \citenamefont {He}}]{he_topological_2019}%
  \BibitemOpen
  \bibfield  {author} {\bibinfo {author} {\bibfnamefont {M.}~\bibnamefont {He}}, \bibinfo {author} {\bibfnamefont {H.}~\bibnamefont {Sun}},\ and\ \bibinfo {author} {\bibfnamefont {Q.~L.}\ \bibnamefont {He}},\ }\bibfield  {title} {\bibinfo {title} {Topological insulator: Spintronics and quantum computations},\ }\bibfield  {journal} {\bibinfo  {journal} {Frontiers of Physics}\ }\textbf {\bibinfo {volume} {14}},\ \href {https://doi.org/10.1007/s11467-019-0893-4} {10.1007/s11467-019-0893-4} (\bibinfo {year} {2019})\BibitemShut {NoStop}%
\bibitem [{\citenamefont {Qin}\ \emph {et~al.}(2023)\citenamefont {Qin}, \citenamefont {Sun}, \citenamefont {Zhou}, \citenamefont {Guo}, \citenamefont {Chen}, \citenamefont {Ke}, \citenamefont {Mao}, \citenamefont {Chen}, \citenamefont {Shao},\ and\ \citenamefont {Zhao}}]{qin_memristor_2023}%
  \BibitemOpen
  \bibfield  {author} {\bibinfo {author} {\bibfnamefont {J.}~\bibnamefont {Qin}}, \bibinfo {author} {\bibfnamefont {B.}~\bibnamefont {Sun}}, \bibinfo {author} {\bibfnamefont {G.}~\bibnamefont {Zhou}}, \bibinfo {author} {\bibfnamefont {T.}~\bibnamefont {Guo}}, \bibinfo {author} {\bibfnamefont {Y.}~\bibnamefont {Chen}}, \bibinfo {author} {\bibfnamefont {C.}~\bibnamefont {Ke}}, \bibinfo {author} {\bibfnamefont {S.}~\bibnamefont {Mao}}, \bibinfo {author} {\bibfnamefont {X.}~\bibnamefont {Chen}}, \bibinfo {author} {\bibfnamefont {J.}~\bibnamefont {Shao}},\ and\ \bibinfo {author} {\bibfnamefont {Y.}~\bibnamefont {Zhao}},\ }\bibfield  {title} {\bibinfo {title} {From spintronic memristors to quantum computing},\ }\href {https://doi.org/10.1021/acsmaterialslett.3c00088} {\bibfield  {journal} {\bibinfo  {journal} {ACS Materials Letters}\ }\textbf {\bibinfo {volume} {5}},\ \bibinfo {pages} {2197} (\bibinfo {year} {2023})}\BibitemShut {NoStop}%
\bibitem [{\citenamefont {Hu}\ \emph {et~al.}(2025)\citenamefont {Hu}, \citenamefont {Huang}, \citenamefont {Cai}, \citenamefont {Wang}, \citenamefont {Yang}, \citenamefont {Cao}, \citenamefont {Xue}, \citenamefont {Huang},\ and\ \citenamefont {He}}]{hu_single_2025}%
  \BibitemOpen
  \bibfield  {author} {\bibinfo {author} {\bibfnamefont {G.}~\bibnamefont {Hu}}, \bibinfo {author} {\bibfnamefont {W.~W.}\ \bibnamefont {Huang}}, \bibinfo {author} {\bibfnamefont {R.}~\bibnamefont {Cai}}, \bibinfo {author} {\bibfnamefont {L.}~\bibnamefont {Wang}}, \bibinfo {author} {\bibfnamefont {C.~H.}\ \bibnamefont {Yang}}, \bibinfo {author} {\bibfnamefont {G.}~\bibnamefont {Cao}}, \bibinfo {author} {\bibfnamefont {X.}~\bibnamefont {Xue}}, \bibinfo {author} {\bibfnamefont {P.}~\bibnamefont {Huang}},\ and\ \bibinfo {author} {\bibfnamefont {Y.}~\bibnamefont {He}},\ }\bibfield  {title} {\bibinfo {title} {Single-electron spin qubits in silicon for quantum computing},\ }\bibfield  {journal} {\bibinfo  {journal} {Intelligent Computing}\ }\textbf {\bibinfo {volume} {4}},\ \href {https://doi.org/10.34133/icomputing.0115} {10.34133/icomputing.0115} (\bibinfo {year} {2025})\BibitemShut {NoStop}%
\bibitem [{\citenamefont {Baibich}\ \emph {et~al.}(1988)\citenamefont {Baibich}, \citenamefont {Broto}, \citenamefont {Fert}, \citenamefont {Van~Dau}, \citenamefont {Petroff}, \citenamefont {Etienne}, \citenamefont {Creuzet}, \citenamefont {Friederich},\ and\ \citenamefont {Chazelas}}]{baibich_giant_1988}%
  \BibitemOpen
  \bibfield  {author} {\bibinfo {author} {\bibfnamefont {M.~N.}\ \bibnamefont {Baibich}}, \bibinfo {author} {\bibfnamefont {J.~M.}\ \bibnamefont {Broto}}, \bibinfo {author} {\bibfnamefont {A.}~\bibnamefont {Fert}}, \bibinfo {author} {\bibfnamefont {F.~N.}\ \bibnamefont {Van~Dau}}, \bibinfo {author} {\bibfnamefont {F.}~\bibnamefont {Petroff}}, \bibinfo {author} {\bibfnamefont {P.}~\bibnamefont {Etienne}}, \bibinfo {author} {\bibfnamefont {G.}~\bibnamefont {Creuzet}}, \bibinfo {author} {\bibfnamefont {A.}~\bibnamefont {Friederich}},\ and\ \bibinfo {author} {\bibfnamefont {J.}~\bibnamefont {Chazelas}},\ }\bibfield  {title} {\bibinfo {title} {Giant {Magnetoresistance} of (001){Fe}/(001){Cr} {Magnetic} {Superlattices}},\ }\href {https://doi.org/10.1103/PhysRevLett.61.2472} {\bibfield  {journal} {\bibinfo  {journal} {Phys. Rev. Lett.}\ }\textbf {\bibinfo {volume} {61}},\ \bibinfo {pages} {2472} (\bibinfo {year} {1988})},\ \bibinfo {note} {copyright (C) 2008 The American Physical Society; Please report any problems to prola@aps.org}\BibitemShut {NoStop}%
\bibitem [{\citenamefont {Slonczewski}(1996)}]{slonczewski_current-driven_1996}%
  \BibitemOpen
  \bibfield  {author} {\bibinfo {author} {\bibfnamefont {J.~C.}\ \bibnamefont {Slonczewski}},\ }\bibfield  {title} {\bibinfo {title} {Current-driven excitation of magnetic multilayers},\ }\href {https://doi.org/10.1016/0304-8853(96)00062-5} {\bibfield  {journal} {\bibinfo  {journal} {Journal of Magnetism and Magnetic Materials}\ }\textbf {\bibinfo {volume} {159}},\ \bibinfo {pages} {L1} (\bibinfo {year} {1996})}\BibitemShut {NoStop}%
\bibitem [{\citenamefont {Berger}(1996)}]{berger_emission_1996}%
  \BibitemOpen
  \bibfield  {author} {\bibinfo {author} {\bibfnamefont {L.}~\bibnamefont {Berger}},\ }\bibfield  {title} {\bibinfo {title} {Emission of spin waves by a magnetic multilayer traversed by a current},\ }\href {https://doi.org/10.1103/PhysRevB.54.9353} {\bibfield  {journal} {\bibinfo  {journal} {Phys. Rev. B}\ }\textbf {\bibinfo {volume} {54}},\ \bibinfo {pages} {9353} (\bibinfo {year} {1996})},\ \bibinfo {note} {copyright (C) 2008 The American Physical Society; Please report any problems to prola@aps.org}\BibitemShut {NoStop}%
\bibitem [{\citenamefont {Sinova}\ \emph {et~al.}(2004)\citenamefont {Sinova}, \citenamefont {Culcer}, \citenamefont {Niu}, \citenamefont {Sinitsyn}, \citenamefont {Jungwirth},\ and\ \citenamefont {MacDonald}}]{sinova_universal_2004}%
  \BibitemOpen
  \bibfield  {author} {\bibinfo {author} {\bibfnamefont {J.}~\bibnamefont {Sinova}}, \bibinfo {author} {\bibfnamefont {D.}~\bibnamefont {Culcer}}, \bibinfo {author} {\bibfnamefont {Q.}~\bibnamefont {Niu}}, \bibinfo {author} {\bibfnamefont {N.~A.}\ \bibnamefont {Sinitsyn}}, \bibinfo {author} {\bibfnamefont {T.}~\bibnamefont {Jungwirth}},\ and\ \bibinfo {author} {\bibfnamefont {A.~H.}\ \bibnamefont {MacDonald}},\ }\bibfield  {title} {\bibinfo {title} {Universal {Intrinsic} {Spin} {Hall} {Effect}},\ }\href {https://doi.org/10.1103/PhysRevLett.92.126603} {\bibfield  {journal} {\bibinfo  {journal} {Phys. Rev. Lett.}\ }\textbf {\bibinfo {volume} {92}},\ \bibinfo {pages} {126603} (\bibinfo {year} {2004})}\BibitemShut {NoStop}%
\bibitem [{\citenamefont {Tserkovnyak}\ \emph {et~al.}(2002)\citenamefont {Tserkovnyak}, \citenamefont {Brataas},\ and\ \citenamefont {Bauer}}]{tserkovnyak_enhanced_2002}%
  \BibitemOpen
  \bibfield  {author} {\bibinfo {author} {\bibfnamefont {Y.}~\bibnamefont {Tserkovnyak}}, \bibinfo {author} {\bibfnamefont {A.}~\bibnamefont {Brataas}},\ and\ \bibinfo {author} {\bibfnamefont {G.~E.~W.}\ \bibnamefont {Bauer}},\ }\bibfield  {title} {\bibinfo {title} {Enhanced {Gilbert} {Damping} in {Thin} {Ferromagnetic} {Films}},\ }\href {https://doi.org/10.1103/PhysRevLett.88.117601} {\bibfield  {journal} {\bibinfo  {journal} {Phys. Rev. Lett.}\ }\textbf {\bibinfo {volume} {88}},\ \bibinfo {pages} {117601} (\bibinfo {year} {2002})},\ \bibinfo {note} {copyright (C) 2008 The American Physical Society; Please report any problems to prola@aps.org}\BibitemShut {NoStop}%
\bibitem [{\citenamefont {Chiba}\ \emph {et~al.}(2014)\citenamefont {Chiba}, \citenamefont {Bauer},\ and\ \citenamefont {Takahashi}}]{chiba_current_2014}%
  \BibitemOpen
  \bibfield  {author} {\bibinfo {author} {\bibfnamefont {T.}~\bibnamefont {Chiba}}, \bibinfo {author} {\bibfnamefont {G.~E.~W.}\ \bibnamefont {Bauer}},\ and\ \bibinfo {author} {\bibfnamefont {S.}~\bibnamefont {Takahashi}},\ }\bibfield  {title} {\bibinfo {title} {Current-induced spin-torque resonance of magnetic insulators},\ }\href {https://doi.org/10.1103/PhysRevApplied.2.034003} {\bibfield  {journal} {\bibinfo  {journal} {Physical Review Applied}\ }\textbf {\bibinfo {volume} {2}},\ \bibinfo {pages} {034003} (\bibinfo {year} {2014})}\BibitemShut {NoStop}%
\bibitem [{\citenamefont {Li}\ \emph {et~al.}(2016)\citenamefont {Li}, \citenamefont {Shelford}, \citenamefont {Shafer}, \citenamefont {Tan}, \citenamefont {Deng}, \citenamefont {Keatley}, \citenamefont {Hwang}, \citenamefont {Arenholz}, \citenamefont {van~der Laan}, \citenamefont {Hicken},\ and\ \citenamefont {Qiu}}]{li_ac_2016}%
  \BibitemOpen
  \bibfield  {author} {\bibinfo {author} {\bibfnamefont {J.}~\bibnamefont {Li}}, \bibinfo {author} {\bibfnamefont {L.~R.}\ \bibnamefont {Shelford}}, \bibinfo {author} {\bibfnamefont {P.}~\bibnamefont {Shafer}}, \bibinfo {author} {\bibfnamefont {A.}~\bibnamefont {Tan}}, \bibinfo {author} {\bibfnamefont {J.~X.}\ \bibnamefont {Deng}}, \bibinfo {author} {\bibfnamefont {P.~S.}\ \bibnamefont {Keatley}}, \bibinfo {author} {\bibfnamefont {C.}~\bibnamefont {Hwang}}, \bibinfo {author} {\bibfnamefont {E.}~\bibnamefont {Arenholz}}, \bibinfo {author} {\bibfnamefont {G.}~\bibnamefont {van~der Laan}}, \bibinfo {author} {\bibfnamefont {R.~J.}\ \bibnamefont {Hicken}},\ and\ \bibinfo {author} {\bibfnamefont {Z.~Q.}\ \bibnamefont {Qiu}},\ }\bibfield  {title} {\bibinfo {title} {Direct detection of pure ac spin current by x-ray pump-probe measurements},\ }\bibfield  {journal} {\bibinfo  {journal} {Physical Review Letters}\ }\textbf {\bibinfo {volume} {117}},\ \href {https://doi.org/10.1103/PhysRevLett.117.076602} {10.1103/PhysRevLett.117.076602} (\bibinfo {year} {2016})\BibitemShut {NoStop}%
\bibitem [{\citenamefont {Kapelrud}\ and\ \citenamefont {Brataas}(2017)}]{kapelrud_pumping_2017}%
  \BibitemOpen
  \bibfield  {author} {\bibinfo {author} {\bibfnamefont {A.}~\bibnamefont {Kapelrud}}\ and\ \bibinfo {author} {\bibfnamefont {A.}~\bibnamefont {Brataas}},\ }\bibfield  {title} {\bibinfo {title} {Spin pumping, dissipation, and direct and alternating inverse spin hall effects in magnetic-insulator/normal-metal bilayers},\ }\bibfield  {journal} {\bibinfo  {journal} {Physical Review B}\ }\textbf {\bibinfo {volume} {95}},\ \href {https://doi.org/10.1103/PhysRevB.95.214413} {10.1103/PhysRevB.95.214413} (\bibinfo {year} {2017})\BibitemShut {NoStop}%
\bibitem [{\citenamefont {Kampfrath}\ \emph {et~al.}(2013)\citenamefont {Kampfrath}, \citenamefont {Battiato}, \citenamefont {Maldonado}, \citenamefont {Eilers}, \citenamefont {Noetzold}, \citenamefont {Maehrlein}, \citenamefont {Zbarsky}, \citenamefont {Freimuth}, \citenamefont {Mokrousov}, \citenamefont {Bluegel}, \citenamefont {Wolf}, \citenamefont {Radu}, \citenamefont {Oppeneer},\ and\ \citenamefont {Muenzenberg}}]{kampfrath_THz_2013}%
  \BibitemOpen
  \bibfield  {author} {\bibinfo {author} {\bibfnamefont {T.}~\bibnamefont {Kampfrath}}, \bibinfo {author} {\bibfnamefont {M.}~\bibnamefont {Battiato}}, \bibinfo {author} {\bibfnamefont {P.}~\bibnamefont {Maldonado}}, \bibinfo {author} {\bibfnamefont {G.}~\bibnamefont {Eilers}}, \bibinfo {author} {\bibfnamefont {J.}~\bibnamefont {Noetzold}}, \bibinfo {author} {\bibfnamefont {S.}~\bibnamefont {Maehrlein}}, \bibinfo {author} {\bibfnamefont {V.}~\bibnamefont {Zbarsky}}, \bibinfo {author} {\bibfnamefont {F.}~\bibnamefont {Freimuth}}, \bibinfo {author} {\bibfnamefont {Y.}~\bibnamefont {Mokrousov}}, \bibinfo {author} {\bibfnamefont {S.}~\bibnamefont {Bluegel}}, \bibinfo {author} {\bibfnamefont {M.}~\bibnamefont {Wolf}}, \bibinfo {author} {\bibfnamefont {I.}~\bibnamefont {Radu}}, \bibinfo {author} {\bibfnamefont {P.~M.}\ \bibnamefont {Oppeneer}},\ and\ \bibinfo {author} {\bibfnamefont {M.}~\bibnamefont {Muenzenberg}},\ }\bibfield  {title} {\bibinfo {title} {Terahertz spin current pulses controlled by magnetic heterostructures},\ }\href {https://doi.org/10.1038/NNANO.2013.43} {\bibfield  {journal} {\bibinfo  {journal} {Nature Nanotechnology}\ }\textbf {\bibinfo {volume} {8}},\ \bibinfo {pages} {256} (\bibinfo {year} {2013})}\BibitemShut {NoStop}%
\bibitem [{\citenamefont {Droegeler}\ \emph {et~al.}(2016)\citenamefont {Droegeler}, \citenamefont {Franzen}, \citenamefont {Volmer}, \citenamefont {Pohlmann}, \citenamefont {Banszerus}, \citenamefont {Wolter}, \citenamefont {Watanabe}, \citenamefont {Taniguchi}, \citenamefont {Stampfer},\ and\ \citenamefont {Beschoten}}]{droegeler_lifetime_2016}%
  \BibitemOpen
  \bibfield  {author} {\bibinfo {author} {\bibfnamefont {M.}~\bibnamefont {Droegeler}}, \bibinfo {author} {\bibfnamefont {C.}~\bibnamefont {Franzen}}, \bibinfo {author} {\bibfnamefont {F.}~\bibnamefont {Volmer}}, \bibinfo {author} {\bibfnamefont {T.}~\bibnamefont {Pohlmann}}, \bibinfo {author} {\bibfnamefont {L.}~\bibnamefont {Banszerus}}, \bibinfo {author} {\bibfnamefont {M.}~\bibnamefont {Wolter}}, \bibinfo {author} {\bibfnamefont {K.}~\bibnamefont {Watanabe}}, \bibinfo {author} {\bibfnamefont {T.}~\bibnamefont {Taniguchi}}, \bibinfo {author} {\bibfnamefont {C.}~\bibnamefont {Stampfer}},\ and\ \bibinfo {author} {\bibfnamefont {B.}~\bibnamefont {Beschoten}},\ }\bibfield  {title} {\bibinfo {title} {Spin lifetimes exceeding 12 ns in graphene nonlocal spin valve devices},\ }\href {https://doi.org/10.1021/acs.nanolett.6b00497} {\bibfield  {journal} {\bibinfo  {journal} {NANO LETTERS}\ }\textbf {\bibinfo {volume} {16}},\ \bibinfo {pages} {3533} (\bibinfo {year} {2016})}\BibitemShut {NoStop}%
\bibitem [{\citenamefont {Xu}\ \emph {et~al.}(2020)\citenamefont {Xu}, \citenamefont {Habib}, \citenamefont {Kumar}, \citenamefont {Wu}, \citenamefont {Sundararaman},\ and\ \citenamefont {Ping}}]{xu_spin_2020}%
  \BibitemOpen
  \bibfield  {author} {\bibinfo {author} {\bibfnamefont {J.}~\bibnamefont {Xu}}, \bibinfo {author} {\bibfnamefont {A.}~\bibnamefont {Habib}}, \bibinfo {author} {\bibfnamefont {S.}~\bibnamefont {Kumar}}, \bibinfo {author} {\bibfnamefont {F.}~\bibnamefont {Wu}}, \bibinfo {author} {\bibfnamefont {R.}~\bibnamefont {Sundararaman}},\ and\ \bibinfo {author} {\bibfnamefont {Y.}~\bibnamefont {Ping}},\ }\bibfield  {title} {\bibinfo {title} {Spin-phonon relaxation from a universal ab initio density-matrix approach},\ }\bibfield  {journal} {\bibinfo  {journal} {Nature Communications}\ }\textbf {\bibinfo {volume} {11}},\ \href {https://doi.org/10.1038/s41467-020-16063-5} {10.1038/s41467-020-16063-5} (\bibinfo {year} {2020})\BibitemShut {NoStop}%
\bibitem [{\citenamefont {Lu}\ \emph {et~al.}(2024)\citenamefont {Lu}, \citenamefont {Wang}, \citenamefont {Zhou}, \citenamefont {Xie}, \citenamefont {Yuan}, \citenamefont {Huang},\ and\ \citenamefont {Yeow}}]{lu_THz_2024}%
  \BibitemOpen
  \bibfield  {author} {\bibinfo {author} {\bibfnamefont {G.}~\bibnamefont {Lu}}, \bibinfo {author} {\bibfnamefont {J.}~\bibnamefont {Wang}}, \bibinfo {author} {\bibfnamefont {R.}~\bibnamefont {Zhou}}, \bibinfo {author} {\bibfnamefont {Z.}~\bibnamefont {Xie}}, \bibinfo {author} {\bibfnamefont {Y.}~\bibnamefont {Yuan}}, \bibinfo {author} {\bibfnamefont {L.}~\bibnamefont {Huang}},\ and\ \bibinfo {author} {\bibfnamefont {J.~T.~W.}\ \bibnamefont {Yeow}},\ }\bibfield  {title} {\bibinfo {title} {Terahertz communication: detection and signal processing},\ }\bibfield  {journal} {\bibinfo  {journal} {Nanotechnology}\ }\textbf {\bibinfo {volume} {35}},\ \href {https://doi.org/10.1088/1361-6528/ad4dad} {10.1088/1361-6528/ad4dad} (\bibinfo {year} {2024})\BibitemShut {NoStop}%
\bibitem [{\citenamefont {Bhide}\ \emph {et~al.}(2024)\citenamefont {Bhide}, \citenamefont {Shetty},\ and\ \citenamefont {Mikkili}}]{bhide_6g_2024}%
  \BibitemOpen
  \bibfield  {author} {\bibinfo {author} {\bibfnamefont {P.}~\bibnamefont {Bhide}}, \bibinfo {author} {\bibfnamefont {D.}~\bibnamefont {Shetty}},\ and\ \bibinfo {author} {\bibfnamefont {S.}~\bibnamefont {Mikkili}},\ }\bibfield  {title} {\bibinfo {title} {Review on 6g communication and its architecture, technologies included, challenges, security challenges and requirements, applications, with respect to ai domain},\ }\bibfield  {journal} {\bibinfo  {journal} {IET Quantum Communication}\ }\href {https://doi.org/10.1049/qtc2.12114} {10.1049/qtc2.12114} (\bibinfo {year} {2024})\BibitemShut {NoStop}%
\bibitem [{\citenamefont {Stiles}\ \emph {et~al.}(2004)\citenamefont {Stiles}, \citenamefont {Xiao},\ and\ \citenamefont {Zangwill}}]{stiles_phenomenological_2004}%
  \BibitemOpen
  \bibfield  {author} {\bibinfo {author} {\bibfnamefont {M.~D.}\ \bibnamefont {Stiles}}, \bibinfo {author} {\bibfnamefont {J.}~\bibnamefont {Xiao}},\ and\ \bibinfo {author} {\bibfnamefont {A.}~\bibnamefont {Zangwill}},\ }\bibfield  {title} {\bibinfo {title} {Phenomenological theory of current-induced magnetization precession},\ }\href {https://doi.org/10.1103/PhysRevB.69.054408} {\bibfield  {journal} {\bibinfo  {journal} {Phys. Rev. B}\ }\textbf {\bibinfo {volume} {69}},\ \bibinfo {pages} {054408} (\bibinfo {year} {2004})},\ \bibinfo {note} {copyright (C) 2008 The American Physical Society; Please report any problems to prola@aps.org}\BibitemShut {NoStop}%
\bibitem [{\citenamefont {Battiato}\ \emph {et~al.}(2010)\citenamefont {Battiato}, \citenamefont {Carva},\ and\ \citenamefont {Oppeneer}}]{battiato_superdiffusion_2010}%
  \BibitemOpen
  \bibfield  {author} {\bibinfo {author} {\bibfnamefont {M.}~\bibnamefont {Battiato}}, \bibinfo {author} {\bibfnamefont {K.}~\bibnamefont {Carva}},\ and\ \bibinfo {author} {\bibfnamefont {P.~M.}\ \bibnamefont {Oppeneer}},\ }\bibfield  {title} {\bibinfo {title} {Superdiffusive spin transport as a mechanism of ultrafast demagnetization},\ }\bibfield  {journal} {\bibinfo  {journal} {Physical Review Letters}\ }\textbf {\bibinfo {volume} {105}},\ \href {https://doi.org/10.1103/PhysRevLett.105.027203} {10.1103/PhysRevLett.105.027203} (\bibinfo {year} {2010})\BibitemShut {NoStop}%
\bibitem [{\citenamefont {van Son}\ \emph {et~al.}(1987)\citenamefont {van Son}, \citenamefont {van Kempen},\ and\ \citenamefont {Wyder}}]{van_son_boundary_1987}%
  \BibitemOpen
  \bibfield  {author} {\bibinfo {author} {\bibfnamefont {P.~C.}\ \bibnamefont {van Son}}, \bibinfo {author} {\bibfnamefont {H.}~\bibnamefont {van Kempen}},\ and\ \bibinfo {author} {\bibfnamefont {P.}~\bibnamefont {Wyder}},\ }\bibfield  {title} {\bibinfo {title} {Boundary {Resistance} of the {Ferromagnetic}-{Nonferromagnetic} {Metal} {Interface}},\ }\href {https://doi.org/10.1103/PhysRevLett.58.2271} {\bibfield  {journal} {\bibinfo  {journal} {Phys. Rev. Lett.}\ }\textbf {\bibinfo {volume} {58}},\ \bibinfo {pages} {2271} (\bibinfo {year} {1987})},\ \bibinfo {note} {copyright (C) 2009 The American Physical Society; Please report any problems to prola@aps.org}\BibitemShut {NoStop}%
\bibitem [{\citenamefont {Stiles}\ and\ \citenamefont {Zangwill}(2002{\natexlab{a}})}]{stiles_noncollinear_2002}%
  \BibitemOpen
  \bibfield  {author} {\bibinfo {author} {\bibfnamefont {M.~D.}\ \bibnamefont {Stiles}}\ and\ \bibinfo {author} {\bibfnamefont {A.}~\bibnamefont {Zangwill}},\ }\bibfield  {title} {\bibinfo {title} {Noncollinear spin transfer in {Co}/{Cu}/{Co} multilayers (invited)},\ }\href {https://doi.org/10.1063/1.1446123} {\bibfield  {journal} {\bibinfo  {journal} {J. Appl. Phys.}\ }\textbf {\bibinfo {volume} {91}},\ \bibinfo {pages} {6812} (\bibinfo {year} {2002}{\natexlab{a}})}\BibitemShut {NoStop}%
\bibitem [{\citenamefont {Stiles}\ and\ \citenamefont {Zangwill}(2002{\natexlab{b}})}]{stiles_anatomy_2002}%
  \BibitemOpen
  \bibfield  {author} {\bibinfo {author} {\bibfnamefont {M.~D.}\ \bibnamefont {Stiles}}\ and\ \bibinfo {author} {\bibfnamefont {A.}~\bibnamefont {Zangwill}},\ }\bibfield  {title} {\bibinfo {title} {Anatomy of spin-transfer torque},\ }\href {https://doi.org/10.1103/PhysRevB.66.014407} {\bibfield  {journal} {\bibinfo  {journal} {Phys. Rev. B}\ }\textbf {\bibinfo {volume} {66}},\ \bibinfo {pages} {014407} (\bibinfo {year} {2002}{\natexlab{b}})},\ \bibinfo {note} {copyright (C) 2008 The American Physical Society; Please report any problems to prola@aps.org}\BibitemShut {NoStop}%
\bibitem [{\citenamefont {Zhou}\ \emph {et~al.}(2023)\citenamefont {Zhou}, \citenamefont {Chen}, \citenamefont {Xu}, \citenamefont {Hou}, \citenamefont {Lai},\ and\ \citenamefont {Yao}}]{zhou_cdx_2023}%
  \BibitemOpen
  \bibfield  {author} {\bibinfo {author} {\bibfnamefont {Q.}~\bibnamefont {Zhou}}, \bibinfo {author} {\bibfnamefont {X.}~\bibnamefont {Chen}}, \bibinfo {author} {\bibfnamefont {X.}~\bibnamefont {Xu}}, \bibinfo {author} {\bibfnamefont {Y.}~\bibnamefont {Hou}}, \bibinfo {author} {\bibfnamefont {T.}~\bibnamefont {Lai}},\ and\ \bibinfo {author} {\bibfnamefont {D.-X.}\ \bibnamefont {Yao}},\ }\bibfield  {title} {\bibinfo {title} {Electronic properties of graphene/cdx (x=s, se, and te) semiconductor heterostructure and a proposal of all-optical injection and detection of electron spins in graphene},\ }\bibfield  {journal} {\bibinfo  {journal} {PHYSICA E-LOW-DIMENSIONAL SYSTEMS \& NANOSTRUCTURES}\ }\textbf {\bibinfo {volume} {146}},\ \href {https://doi.org/10.1016/j.physe.2022.115559} {10.1016/j.physe.2022.115559} (\bibinfo {year} {2023})\BibitemShut {NoStop}%
\bibitem [{\citenamefont {Bass}\ and\ \citenamefont {Pratt}(2007)}]{bass_spin-diffusion_2007}%
  \BibitemOpen
  \bibfield  {author} {\bibinfo {author} {\bibfnamefont {J.}~\bibnamefont {Bass}}\ and\ \bibinfo {author} {\bibfnamefont {W.~P.}\ \bibnamefont {Pratt}},\ }\bibfield  {title} {\bibinfo {title} {Spin-diffusion lengths in metals and alloys, and spin-flipping at metal/metal interfaces: an experimentalist's critical review},\ }\href {https://doi.org/10.1088/0953-8984/19/18/183201} {\bibfield  {journal} {\bibinfo  {journal} {Journal of Physics: Condensed Matter}\ }\textbf {\bibinfo {volume} {19}},\ \bibinfo {pages} {183201} (\bibinfo {year} {2007})}\BibitemShut {NoStop}%
\bibitem [{\citenamefont {Novoselov}\ \emph {et~al.}(2004)\citenamefont {Novoselov}, \citenamefont {Geim}, \citenamefont {Morozov}, \citenamefont {Jiang}, \citenamefont {Zhang}, \citenamefont {Dubonos}, \citenamefont {Grigorieva},\ and\ \citenamefont {Firsov}}]{novoselov_2d_2004}%
  \BibitemOpen
  \bibfield  {author} {\bibinfo {author} {\bibfnamefont {K.}~\bibnamefont {Novoselov}}, \bibinfo {author} {\bibfnamefont {A.}~\bibnamefont {Geim}}, \bibinfo {author} {\bibfnamefont {S.}~\bibnamefont {Morozov}}, \bibinfo {author} {\bibfnamefont {D.}~\bibnamefont {Jiang}}, \bibinfo {author} {\bibfnamefont {Y.}~\bibnamefont {Zhang}}, \bibinfo {author} {\bibfnamefont {S.}~\bibnamefont {Dubonos}}, \bibinfo {author} {\bibfnamefont {I.}~\bibnamefont {Grigorieva}},\ and\ \bibinfo {author} {\bibfnamefont {A.}~\bibnamefont {Firsov}},\ }\bibfield  {title} {\bibinfo {title} {Electric field effect in atomically thin carbon films},\ }\href@noop {} {\bibfield  {journal} {\bibinfo  {journal} {Science}\ }\textbf {\bibinfo {volume} {306}},\ \bibinfo {pages} {666} (\bibinfo {year} {2004})}\BibitemShut {NoStop}%
\bibitem [{\citenamefont {Xue}\ \emph {et~al.}(2021)\citenamefont {Xue}, \citenamefont {Zhong},\ and\ \citenamefont {Ma}}]{xue_graphene_2021}%
  \BibitemOpen
  \bibfield  {author} {\bibinfo {author} {\bibfnamefont {Z.}~\bibnamefont {Xue}}, \bibinfo {author} {\bibfnamefont {S.}~\bibnamefont {Zhong}},\ and\ \bibinfo {author} {\bibfnamefont {Y.}~\bibnamefont {Ma}},\ }\bibfield  {title} {\bibinfo {title} {Graphene-fss hybrid absorptive structure with amplitude/frequency dual-modulated passband},\ }\href {https://doi.org/10.1109/LAWP.2021.3094835} {\bibfield  {journal} {\bibinfo  {journal} {IEEE Antennas and Wireless Propagation Letters}\ }\textbf {\bibinfo {volume} {20}},\ \bibinfo {pages} {1711} (\bibinfo {year} {2021})}\BibitemShut {NoStop}%
\bibitem [{\citenamefont {Cultrera}\ \emph {et~al.}(2024)\citenamefont {Cultrera}, \citenamefont {Serazio}, \citenamefont {Fabricius},\ and\ \citenamefont {Callegaro}}]{cultrera_iec_2024}%
  \BibitemOpen
  \bibfield  {author} {\bibinfo {author} {\bibfnamefont {A.}~\bibnamefont {Cultrera}}, \bibinfo {author} {\bibfnamefont {D.}~\bibnamefont {Serazio}}, \bibinfo {author} {\bibfnamefont {N.}~\bibnamefont {Fabricius}},\ and\ \bibinfo {author} {\bibfnamefont {L.}~\bibnamefont {Callegaro}},\ }\bibfield  {title} {\bibinfo {title} {New iec standards for the measurement of sheet resistance on large-area graphene using the van der pauw and the in-line four-point probe methods},\ }\bibfield  {journal} {\bibinfo  {journal} {Measurement}\ }\textbf {\bibinfo {volume} {236}},\ \href {https://doi.org/10.1016/j.measurement.2024.114980} {10.1016/j.measurement.2024.114980} (\bibinfo {year} {2024})\BibitemShut {NoStop}%
\end{thebibliography}%

\onecolumngrid

\newpage
\appendix 

\section{Parameter Settings for the Ring Resonator}

\begin{figure}[t]
    \includegraphics[width=0.4\linewidth]{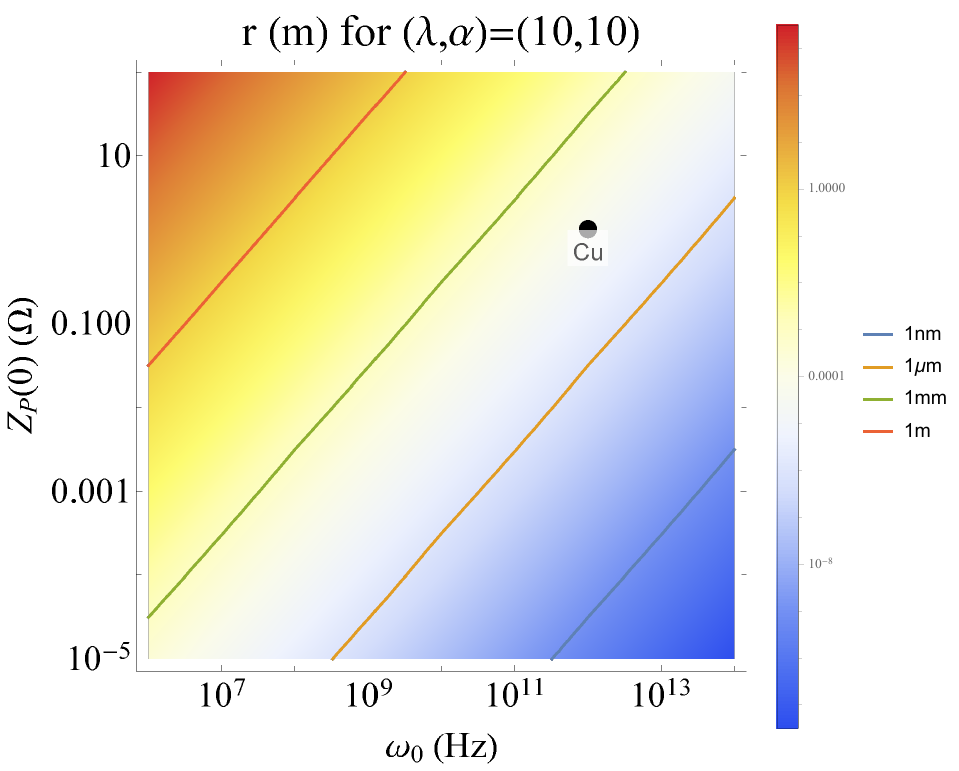}
    \hspace{0.07\linewidth}
    \includegraphics[width=0.4\linewidth]{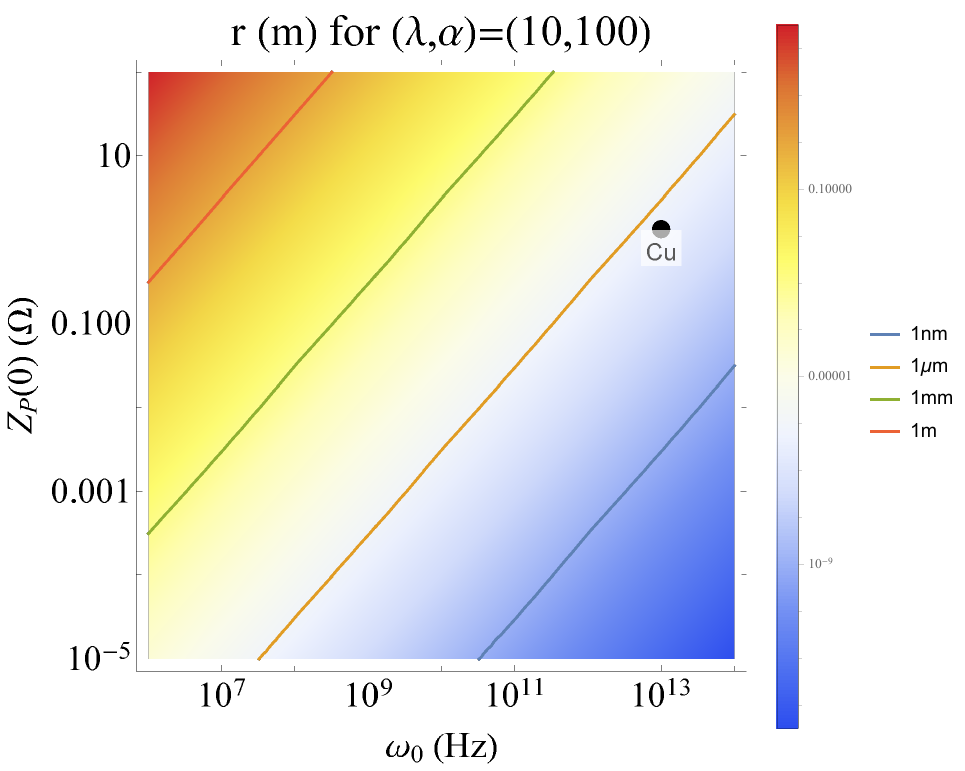}
    \caption{The suggested value of ring radius $r$ for different DC parallel resistance $Z_\ssf{P}(0)$ and resonant frequency $\omega_0$ for spin-valve-based ring oscillators.
    Left: Results with $\lambda=l_s^\ssf{N}/d=10$ and $\alpha=\omega_0\tau_s^\ssf{N}=10$.
    Right: Results with $l_s^\ssf{N}/d=10$ and $\omega_0\tau_s^\ssf{N}=100$.
    The specific parameters and $r$ results for Cu (according to main text) is shown by the black point.}
    \label{fig:para}
\end{figure}

Here we provide an estimation of the parameter settings in the $RLC$ ring resonator. For simplicity, we only consider the capacitance caused by the spin accumulation at the interface, and the inductance contributed by the ring. The (self) inductance of such a ring with radius $r$ is

\begin{equation}
    L\sim\frac{\mu_0 I/2r\cdot \pi r^2}{I}=\frac{\pi}{2}\mu_0 r\sim\mu_0 r.
\end{equation}
For the capacitance, we analyze an antiparallel spin valve with $D_F\rightarrow0$. With $l_s^\ssf{N}/d=\lambda$ and $\omega_0\tau_s^\ssf{N}=\alpha$, the complex wave vector $\kappa_N=\sqrt{1+i\alpha}/l_s^\ssf{N}$, and the boundary impedance will be
\begin{equation}
    Z=Z_\ssf{P}(0)\frac{2\lambda}{\sqrt{1+i\alpha}}\coth\frac{\sqrt{1+i\alpha}}{2\lambda}
    \equiv Z_\ssf{P}(0) f(\lambda,\alpha),
\end{equation}
in which $Z_\ssf{P}(0)=\rho^\ssf{N}d/A$ is the boundary resistance of a parallel spin valve in DC condition, and $A$ is the cross section of the interface. According to \Figure{fig:FNF}, it is advisable to make $\lambda=10$ and $\alpha\gtrsim10$ for $Z$ to approach totally capacitive behavior. Suppose $\abs{Z}\sim 1/\omega_0 C$, the equivalent capacitance is
\begin{equation}
    C=\frac{1}{\omega_0\abs{Z}}=\frac{1}{\omega_0 Z_\ssf{P}(0)\abs{f(\lambda,\alpha)}}.
\end{equation}
For the ring resonator with characteristic frequency $\omega_0=1/\sqrt{LC}$, the radius $r$ should satisfy the following relation:
\begin{equation}
    \omega_0^2\cdot\mu_0r\cdot\frac{1}{\omega_0 Z_\ssf{P}(0)\abs{f(\lambda,\alpha)}}\sim1\Longrightarrow r\sim\frac{Z_\ssf{P}(0)\abs{f(\lambda,\alpha)}}{\mu_0\omega_0}.
\end{equation}
The predicted values of $r$ are illustrated in \Figure{fig:para}, in which $Z_\ssf{P}(0)$ and $\omega_0$ are chosen as variables. We show the results for $(\lambda,\alpha)=(10,10)$ and $(10,100)$ for $Z$ to be almost capacitive. The desirable $r$ is positively correlated to $Z_\ssf{P}(0)/\omega_0$. To keep $r$ small, one needs lower impedance for lower frequency. This is consistent with the main text:
\begin{equation}
    \abs{f(10,10)}\sim10\Longrightarrow r\sim10\frac{\rho^\ssf{N}d}{A\mu_0\omega_0}=\frac{\rho^\ssf{N}d\tau_s^\ssf{N}}{A\mu_0}.
\end{equation}

\end{document}